\documentstyle[12pt,epsf]{article} 

\def\mathrm{\rm}

\newcommand {\bea} {\begin{eqnarray}}
\newcommand {\eea} {\end{eqnarray}}
\newcommand {\beq} {\begin{equation}}
\newcommand {\eeq} {\end{equation}}

\begin{document}   
\begin{titlepage}
 \begin{flushright}
  SLAC-PUB-8199\\
  Revised July 1999
 \end{flushright}

\begin{center}
 {\Large\bf Direct Measurement of $A_c$ using Inclusive Charm Tagging
 at the SLD Detector$^*$}
\end{center}

\begin{center}
{\bf The SLD Collaboration$^{**}$}

{\it Stanford Linear Accelerator Center

Stanford University, Stanford, CA 94309 }
\end{center}

\bigskip
\vspace{1.cm} 

\begin{center}
{\bf Contact: Thomas Wright, twright@slac.stanford.edu}
\end{center}

\vspace{1.cm} 

\begin{abstract}
\noindent


We report a new measurement of $A_c$ using data obtained by SLD in
1993-98.  This measurement uses a vertex tag technique, where the
selection of a $c$ hemisphere is based on the reconstructed mass of
the charm hadron decay vertex. The method uses the 3D vertexing
capabilities of SLD's CCD vertex detector and the small and stable SLC
beams to obtain a high $c$-event tagging efficiency and purity of
28\% and 82\%, respectively.  Charged kaons identified by the CRID
detector and the charge of the reconstructed vertex provide an
efficient quark-antiquark tag, with the analyzing power calibrated from the
data.  We obtain a preliminary result
of $A_c = 0.603 \pm 0.028 \pm 0.023$

\end{abstract}

\vfill
 
\begin{center}
{\em Contributed to the International Europhysics Conference on High
Energy Physics, July 15-21 1999, Tampere, Finland; Ref 6-474, and to
the XIXth International Symposium on Lepton Photon Interactions,
August 9-14 1999, Stanford, USA.}
\end{center}

$^*$Work supported by Department of Energy contract
DE--AC03--76SF00515 (SLAC).

\end{titlepage}
\clearpage

\section{Introduction}
 
Measurements of fermion asymmetries at the $Z^0$ resonance probe a
combination of the vector and axial vector couplings of the $Z^0$ to
fermions, $ A_f = 2 v_f a_f / (v_f^2+a_f^2) $.  The parameters $A_{f}$
express the extent of parity violation at the $Zff$ vertex and provide
sensitive tests of the Standard Model.
 
The Born-level differential cross section for the reaction $e^{+}
e^{-} \rightarrow Z^0 \rightarrow f\bar{f}$ is

\begin{equation}
\frac{d \sigma_f}{dz} \propto
(1-A_e P_e) (1+z^2) + 2A_f (A_e - P_e) z \, ,
\end{equation}

where $P_e$ is the longitudinal polarization of the electron beam
($P_e > 0$ for right-handed (R) polarization) and $z = \cos\theta$ is
the direction of the outgoing fermion relative to the incident
electron.  The parameter $A_{f}$ can be isolated by forming the
left-right forward-backward asymmetry ${\tilde{A}}^{f}_{FB}(z) = |P_e|
A_f \, 2z / (1+z^2) \, ,$ although in this analysis we work directly
with the basic cross section.

This note describes the analysis of the data taken during 1996-98 with the
newer VXD3 vertex detector.  Analysis of the 1993-95 data taken with the
original VXD2 vertex detector is described in \cite{ac97}. 

\section{The SLD Detector}
 
The operation of the SLAC Linear Collider with a polarized electron
beam has been described in detail elsewhere \cite{SLC}.  During the
1996 run, the SLC Large Detector (SLD)~\cite{SLD} recorded 100k
events with a luminosity-weighted electron beam
polarization of $ |P_e| = 0.765 \pm 0.005 $. In 1997-98 a sample of
350k events with average polarization of $ |P_e| = 0.733 \pm
0.008 $ was obtained.

Charged particle tracking and momentum analysis are provided by the
Central Drift Chamber \cite{CDC} and the CCD-based vertex detector
\cite{VXD}.  The Liquid Argon Calorimeter (LAC) \cite{LAC} measures
the energy of charged and neutral particles and is also used for
electron identification.  Muon tracking is provided by the Warm Iron
Calorimeter (WIC) \cite{WIC}.  The Cherenkov Ring Imaging Detector
(CRID) \cite{CRID} information (limited to the barrel region) provides
particle identification.  It consists of liquid and gas Cherenkov
radiators illuminating large area UV photon detectors.  Only the gas
information has been included in this analysis, since the liquid
covers only marginally the interesting momentum region.

\section{Event Selection} 

Hadronic events are selected based on the visible energy and track
multiplicity in the event. The visible energy is measured using
central drift chamber (CDC) tracks and must exceed 18 GeV.  There must
be at least 7 CDC tracks, 3 with hits in the vertex
detector.  We also require that the thrust axis, measured from
calorimeter clusters, satisfy $|cos{\theta_{thr}}|<0.7$.  This ensures
that the event is contained within the acceptance of the vertex
detector.  All detector elements are also required to be fully
operational. Additionally, we restrict events to 3 jets or less to
make sure that we have well defined hemispheres.  Jets are defined by
the JADE algorithm~\cite{JADE} with $ycut = 0.02$.  A total of 290k
events pass the above hadronic event selection and jet cut.
Background, predominately due to taus, is estimated at $<0.1$\%.

The SLC interaction point (IP) has a size of approximately $(1.5\times
0.5\times 700)$ ${\mu}$m in ($x$,$y$,$z$). The motion of the IP $xy$
position over a short time interval is estimated to be $\sim6$ ${\mu}$m.
Because this motion is smaller than the $xy$ resolution obtained by
fitting tracks to find the primary vertex (PV) in a given event, we
use the average IP position for the $x$ and $y$ coordinates of the
primary vertex.  This average is obtained from tracks with hits in the
vertex detector in ~30 sequential hadronic events.  The $z$ coordinate
of the PV is determined event-by-event.  This results in a
PV uncertainty of $\sim6$ ${\mu}$m transverse and $\sim25$~${\mu}$m
longitudinal to the beam direction.

\subsection{Track Selection}

Reconstruction of the mass of heavy hadrons is initiated by
identifying secondary vertices in each hemisphere. Only tracks that
are well measured are included in the vertex and mass
reconstruction. Tracks are required to have at least 23 CDC hits and
start within a radius of 50 cm of the IP.  The CDC track is also
required to extrapolate to within 1.0 cm of the IP in $xy$ and within
1.5 cm of the PV in $z$.  At least two vertex detector hits are
required, the combined drift chamber + vertex detector fit must
satisfy $\chi^2/{d.o.f.} < 8$, and $|cos\theta|<0.87$. Tracks with an
$xy$ impact parameter $>3.0$ mm or an $xy$ impact parameter error
$>250$ ${\mu}$m with respect to the IP are removed from consideration in
the vertex reconstruction.

\subsection{Vertex Mass Reconstruction}

Vertex identification is done topologically.~\cite{DJNIM}.  This
method searches for space points in 3D where track density functions
overlap. Each track is parameterized by a Gaussian probability density
tube with a width equal to the uncertainty in the measured track
position at the IP.  Points in space where there is a large overlap of
probability density are considered as possible vertex points.  Final
selection of vertices is done by clustering maxima in the overlap
density distribution into vertices for separate hemispheres.  We found
secondary vertices in 84\% of bottom, 38\% of charm, and 2\% of light quark
events.



The mass of the secondary vertex is calculated using the tracks that
are associated with the vertex.  Each track is assigned the mass of a
charged pion and the invariant mass of the vertex is calculated.  The
reconstructed mass is corrected to account for neutral particles as
follows.  Using kinematic information from the vertex flight path and
the momentum sum of the tracks associated with the secondary vertex,
we add a minimum amount of missing momentum to the invariant mass.
This is done by assuming the true quark momentum is aligned with the
flight direction of the vertex.  The so-called $P_t$-corrected mass is
given by:
$$M_{VTX} = \sqrt{{M^2}_{tk} + {P_t}^2} + |P_t|$$ where $M_{tk}$ is the
mass for the tracks associated with the secondary vertex.  We restrict
the contribution to the invariant mass that the additional transverse
momentum adds to be less than the initial mass of the secondary
vertex. This cut ensures that poorly measured vertices in $uds$ events
do not leak into the sample by adding large $P_t$.

\subsection{Flavor Tag}

A bottom tag is defined as a hemisphere with an invariant mass above 2
GeV$/c^2$. The intermediate mass region, between 0.5 and 2 GeV$/c^2$ contains a
mixture of $b$ and $c$, with a small $uds$ background. We define some
additional cuts to reject $b$ and $uds$. A charm tag is defined as
follows:

\begin{itemize}
\item $0.55 < M_{VTX} < 2$ GeV$/c^2$
\item Vertex momentum ($P_{VTX}$) greater than 5 GeV$/c$.
\item Fragmentation cut: $15M_{VTX} - P_{VTX} < 10$. This uses the fact that
      D hadrons from direct charm have a higher momentum
      than those from $B$ hadron decays.
\end{itemize}

\begin{center}
\begin{figure}[htb]
  \epsfxsize12cm
  \hspace{2cm}
  \epsffile{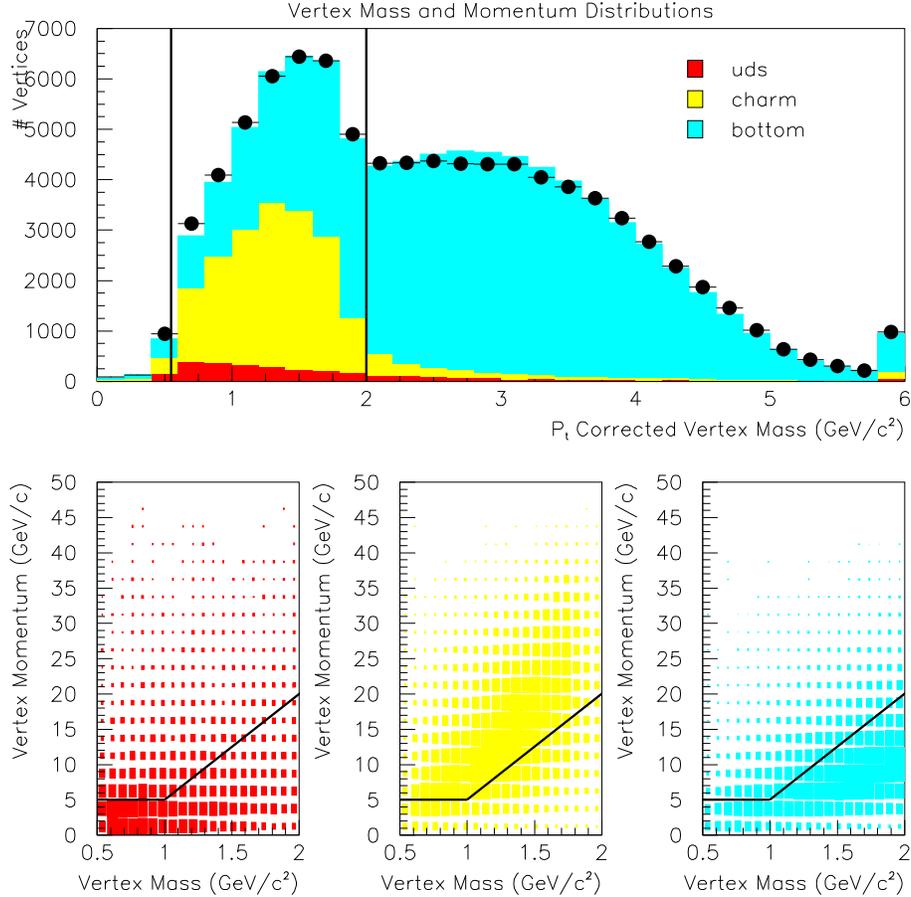}
  \caption{Distributions of $M_{VTX}$ and $P_{VTX}$ vs. $M_{VTX}$ for 
           $uds$, $c$, and $b$ hemispheres.  The lines represent the cuts
           described in the text.  The points on the $M_{VTX}$ plot are 
           the data.}
  \label{fig_mass}
\end{figure}
\end{center}

These tags are calibrated against the data as described in \cite{rc97}.  The
efficiencies $\eta_c$ (for the charm tag) and $\epsilon_b$ (for the bottom
tag), and partial widths $R_c$ and $R_b$ are found by comparing the single-
vs. double-tagged event rates for the two tags.  In addition, the $b$
mis-tag efficiency $\eta_b$ can be found from the fraction of events with
a charm tag in one hemisphere and a bottom tag in the other (mixed tag).  The
light-flavor efficiencies and $\epsilon_c$ are taken from Monte Carlo.

A charm event is defined to be one with at least one charm-tagged
hemisphere and no bottom-tagged hemispheres.  This is found to be
$\sim 28\%$ efficient for charm events.  With the calibrated
efficiencies and partial widths the charm purity of these events is
calculated to be $f_c = 82.1\pm0.5\%$.  This is in good agreement with the
Monte Carlo value 82.6\%.  The $b$ background fraction is
15.5\% with $uds$ making up the remainder.

\subsection{Signal Tag}

The determination of the direction of the quark is done in two ways.
The first is the vertex charge, $Q_{VTX}$.  Charged vertices, coming
mostly from $D^{\pm}$ and $D_s$ have a positive charge for $c$
vertices and a negative charge for $\bar c$.  One would expect that
the $b$ background has an opposite sign, but in reality this is
diluted significantly.  The $b$ vertices that survive the charm tag
usually miss some tracks and therefore have lost most of their
quark-antiquark charge correlation information.

The second method is the kaon charge, $Q_K$. This is the total charge
of the CRID-identified kaon tracks in the vertex. For the kaon charge, the
signals for $b\rightarrow c\rightarrow s$ and $c \rightarrow s$ decays
have the same sign.

\begin{center}
\begin{figure}[htb]
  \epsfxsize12cm
\hspace{2cm}
  \epsffile{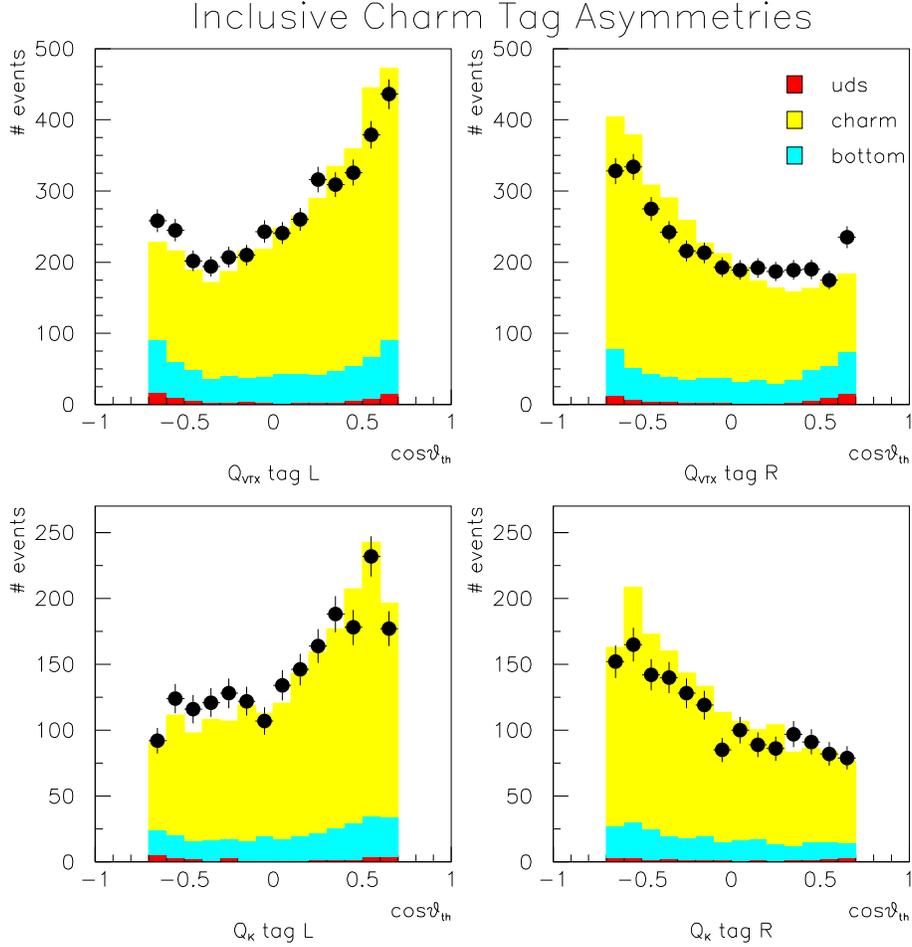}
  \caption{Measured asymmetry in the vertex charge and kaon channel
           for data (points) and MC (solid).  Left side shows left-polarized electrons and similarly for right side.}
  \label{fig_acsig}
\end{figure}
\end{center}

We have $\sim25$\% efficiency for the kaon tag and
$\sim50$\% for the vertex charge for both charm and bottom events.
A hemisphere is considered charged if it has either tag nonzero, but
hemispheres with the two tags in disagreement are considered uncharged.


The probability to correctly discrimate between quark/antiquark for
these tags can be calibrated from the data.  The sample used is the
events with both hemispheres charm-tagged and with nonzero charge in
each hemisphere.  The fraction of these events that are in agreement
(opposite charges) can be written as $r_{agree} = p_{correct}^2 +
(1-p_{correct})^2$.  This simply says the hemispheres must either be
both right or both wrong.  After making a correction for the $b$
contamination we find $p_c^{correct} = 94.2\pm1.2\%$.  The Monte Carlo
gives 92.1\%.

The mixed tag events defined above that also have double charge are
used to calibrate the $b$ background in a similar way.  We find
$p_b^{correct} = 60.5\pm3.6\%$ (the Monte Carlo value is 57.7\%).
Because this procedure calibrates the quark/antiquark flavor at
production the $B$-mixing dilution is automatically included in
$p_b^{correct}$.

\section{Results}

   A maximum likelihood fit of all tagged events is used to
determine $A_{c}$.  As a likelihood function we use the total cross
section:

\begin{eqnarray}
{\cal L} & \propto & (1+z^2)(1-A_e P_e) + 2z(A_e-P_e) \\
& & \left\{ f_c(2p_c^{correct}-1)A_c + \right. \nonumber \\
& & \left. f_b(2p_b^{correct}-1)A_b + f_{uds}A_{uds} \right\} \nonumber
\end{eqnarray}

where $z=-Q\cos\theta_{thr}$, the thrust axis signed by the tagging method
described earlier, is an estimate of the quark direction, $f_{c,b,uds}$ is
the probability for an event to be $c,b$, or $uds$ respectively, and the
factor $(2p^{correct}-1)$ is the effectiveness of the quark/antiquark tag.
The shape of these functions in $z$ and $P_{VTX}$ is taken from Monte Carlo
with the overall normalizations determined from the data. The light-flavor
raw asymmetry $A_{uds}$ is taken from the MC simulation. The three signs
governing the left-right forward-backward asymmetry -- beam polarization
$P_e$, hemisphere tag charge $Q$, and quark direction $\cos\theta_{thr}$ --
are incorporated automatically into the likelihood function.


The QCD corrections to the cross-section are well known~\cite{QCDCOR}.
We account for them with a correction term:

\[ 
A^q_{FB|O(\alpha_s)}(\theta) 
= A^q_{FB|O(0)}(\theta) ( 1-\Delta^q_{O(\alpha_s)}(\theta)) 
\]

These QCD corrections have to be adjusted for any bias in the analysis
method against $q\bar q g$ events:

\[ \Delta_{\mbox{QCD}}^{eff} = f  \Delta_{\mbox{QCD}} \]

We estimated the analysis bias factor $f^{QCD}$ for $b$ and $c$ from a
generator-level Monte Carlo study:

\[ 
f^{QCD} = \frac{A_{q\bar q}^{gen} - A_{q\bar q+q\bar q g}^{analysis}}
{A_{q\bar q}^{gen} - A_{q\bar q+q\bar q g}^{gen}}
\]

We found $f^{QCD}_c = 0.25 \pm 0.06$ and $f^{QCD}_b = 0.31 \pm 0.08$.

From the sample of selected events we measure $A_c = 0.589 \pm 0.031$.
Combining with the previous SLD result for 1993-95 data \cite{ac97} we
obtain:
\[ A_c = 0.603 \pm 0.028 \]

The error is statistical only.

\section{Systematic Errors}

The systematic errors for the 1996-98 SLD result can be found
in Table~\ref{systable}. We give a brief description of the different
sources.

\begin{table}
\caption{Systematic errors for the maximum likelihood analysis}
\label{systable}
\begin{center}
\begin{tabular}{|l|l|} 
\hline
Source & $\delta A_c$ \\ 
\hline
{\bf Tag Composition }&  \\ 
\hline
$f_b$, $f_c$ & 0.002 \\
$uds$ efficiency & 0.004\\
$uds$ asymmetry & 0.001 \\ 
\hline
{\bf Analyzing Power} & \\
\hline
$p_c^{correct}$ & 0.016\\
$p_b^{correct}$ & 0.015\\
Tracking efficiency 3\%& 0.003 \\
MC Statistics & 0.003 \\ 
\hline
{\bf Fit Systematics} & \\ 
\hline
$f,p^{correct}$ shape & 0.006 \\
$\delta P_e$   & 0.007 \\
$A_b$ ($0.935 \pm 0.040$) & 0.002 \\ 
\hline
{\bf QCD corrections} &   \\ 
\hline
analysis bias & 0.002 \\
$\alpha_s (0.118\pm0.007)$ & 0.003 \\
$g \rightarrow c\bar c$ & 0.001 \\ 
\hline
{\bf Total} & 0.025  \\ 
\hline
\end{tabular}
\end{center}
\end{table}

The flavor composition error includes the statistical error associated with
the calibration procedure.  Also included is the uncertainty on the $uds$
efficiency from Monte Carlo. The $uds$ asymmetry was
found to be compatible with zero in the Monte Carlo. We vary by the MC
statistical error of $\pm 0.15$ to estimate the uncertainty from this
source.

The analyzing power error comes mostly from calibration statistics.
Also included here are the tracking efficiency uncertainties and MC
statistics which impact the inter-hemisphere correlation.

The fit systematics include the shape of $f$ and $p$ as functions of
$\cos\theta_{thr}$ and $P_{VTX}$.  These shapes are taken from the
Monte Carlo and normalized by the calibrated values.  The error is
estimated by fitting with and without these shapes.  Also included in
this category are the $b$ quark asymmetry and beam polarization which
are needed in the fit.

The error from QCD corrections comes mainly from the conservative
uncertainty on $\alpha_s$.  Also contributing are the correction
factors $f^{QCD}_{c,b}$ and dilution from events with gluon splitting.

\section{Conclusions} We have performed a measurement of $A_c$ using a
method that takes advantage of some of the unique features of the SLC/SLD
experimental program. Our preliminary result based on 450k hadronic $Z^0$
decays is:

\[ A_c = 0.603 \pm 0.028  \pm 0.023 ~~~~~\mbox{\bf Preliminary} \]

This result is consistent with the SM expectation of 0.67 and other
measurements at SLD and LEP.  Due to the efficient charm tag and high
analyzing power quark-antiquark discrimination, the
statistical power of this analysis is significantly improved compared
to more conventional techniques.  Because the systematic errors are
dominated by calibration statistics this result is largely uncorrelated
with other measurements.

\section*{Acknowledgements}
We thank the personnel of the SLAC accelerator department and the
technical
staffs of our collaborating institutions for their outstanding efforts
on our behalf.

\vskip .5truecm

\vbox{\footnotesize\renewcommand{\baselinestretch}{1}\noindent
$^*$Work supported by Department of Energy
  contracts:
  DE-FG02-91ER40676 (BU),
  DE-FG03-91ER40618 (UCSB),
  DE-FG03-92ER40689 (UCSC),
  DE-FG03-93ER40788 (CSU),
  DE-FG02-91ER40672 (Colorado),
  DE-FG02-91ER40677 (Illinois),
  DE-AC03-76SF00098 (LBL),
  DE-FG02-92ER40715 (Massachusetts),
  DE-FC02-94ER40818 (MIT),
  DE-FG03-96ER40969 (Oregon),
  DE-AC03-76SF00515 (SLAC),
  DE-FG05-91ER40627 (Tennessee),
  DE-FG02-95ER40896 (Wisconsin),
  DE-FG02-92ER40704 (Yale);
  National Science Foundation grants:
  PHY-91-13428 (UCSC),
  PHY-89-21320 (Columbia),
  PHY-92-04239 (Cincinnati),
  PHY-95-10439 (Rutgers),
  PHY-88-19316 (Vanderbilt),
  PHY-92-03212 (Washington);
  The UK Particle Physics and Astronomy Research Council
  (Brunel, Oxford and RAL);
  The Istituto Nazionale di Fisica Nucleare of Italy
  (Bologna, Ferrara, Frascati, Pisa, Padova, Perugia);
  The Japan-US Cooperative Research Project on High Energy Physics
  (Nagoya, Tohoku);
  The Korea Research Foundation (Soongsil, 1997).}

\bibliographystyle{plain}

\begin{thebibliography}{60}

\bibitem{ac97} SLD Collab., K. Abe {\it et al.}, SLAC-PUB 7595, 
   Submitted to Lepton-Photon, Hamburg, Germany, 1997.
\bibitem{SLC} SLD Collab., K. Abe {\it et al.}, Phys. Rev. Lett.
   {\bf 73}, 25 (1994).
\bibitem{SLD} SLD Collab., K. Abe {\it et al.}, Phys. Rev. 
   {\bf D53}, 1023 (1996).
\bibitem{CDC} M. Hildreth {\it et al.}, Nucl. Inst. Meth. 
   {\bf A367}, 111 (1995).
\bibitem{VXD} K. Abe {\it et al.}, Nucl. Inst. Meth.
   {\bf A400}, 287 (1997).
\bibitem{LAC} D. Axen {\it et al.}, Nucl. Inst. Meth.
   {\bf A328}, 472 (1993).
\bibitem{WIC} A. Benvenuti {\it et al.}, Nucl. Inst. Meth.
   {\bf A276}, 94 (1989); {\bf A290}, 353 (1990).
\bibitem{CRID} K. Abe {\it et al.}, Nucl. Inst. Meth.
   {\bf A343}, 74 (1994).
\bibitem{JADE} W. Bartel {\it et el.}, Z. Phys. {\bf C33}, 23 (1986).
\bibitem{DJNIM} D. Jackson, Nucl. Inst. Meth. {\bf A388}, 247 (1997).
\bibitem{rc97} SLD Collab., K. Abe {\it et al.}, SLAC-PUB 7594,
   Submitted to Lepton-Photon, Hamburg, Germany, 1997.
\bibitem{QCDCOR} J. B. Stav and H. A. Olsen, Phys. Rev. 
   {\bf D52}, 1359 (1995); Phys. Rev. {\bf D50}, 6775 (1994).
\end{thebibliography}

%
%
%
\section*{$^{**}$ List of Authors}

\begin{center}
\def\iADEL{$^{(1)}$}
\def\iAOMORI{$^{(2)}$}
\def\iBOLO{$^{(3)}$}
\def\iBRI{$^{(4)}$}
\def\iBRUN{$^{(5)}$}
\def\iBU{$^{(6)}$}
\def\iCINC{$^{(7)}$}
\def\iCOLO{$^{(8)}$}
\def\iCOLU{$^{(9)}$}
\def\iCSU{$^{(10)}$}
\def\iFERR{$^{(11)}$}
\def\iFRAS{$^{(12)}$}
\def\iILLI{$^{(13)}$}
\def\iJHU{$^{(14)}$}
\def\iLBL{$^{(15)}$}
\def\iLTU{$^{(16)}$}
\def\iMASS{$^{(17)}$}
\def\iMISSI{$^{(18)}$}
\def\iMIT{$^{(19)}$}
\def\iMOSCOW{$^{(20)}$}
\def\iNAGO{$^{(21)}$}
\def\iOREG{$^{(22)}$}
\def\iOXF{$^{(23)}$}
\def\iPADO{$^{(24)}$}
\def\iPERU{$^{(25)}$}
\def\iPISA{$^{(26)}$}
\def\iRAL{$^{(27)}$}
\def\iRUTG{$^{(28)}$}
\def\iSLAC{$^{(29)}$}
\def\iSOGA{$^{(30)}$}
\def\iSOONG{$^{(31)}$}
\def\iTENN{$^{(32)}$}
\def\iTOHO{$^{(33)}$}
\def\iUCSB{$^{(34)}$}
\def\iUCSC{$^{(35)}$}
\def\iUVIC{$^{(36)}$}
\def\iVAND{$^{(37)}$}
\def\iWASH{$^{(38)}$}
\def\iWISC{$^{(39)}$}
\def\iYALE{$^{(40)}$}

  \baselineskip=.75\baselineskip  
\mbox{Kenji  Abe\unskip,\iNAGO}
\mbox{Koya Abe\unskip,\iTOHO}
\mbox{T. Abe\unskip,\iSLAC}
\mbox{I. Adam\unskip,\iSLAC}
\mbox{T.  Akagi\unskip,\iSLAC}
\mbox{H. Akimoto\unskip,\iSLAC}
\mbox{N.J. Allen\unskip,\iBRUN}
\mbox{W.W. Ash\unskip,\iSLAC}
\mbox{D. Aston\unskip,\iSLAC}
\mbox{K.G. Baird\unskip,\iMASS}
\mbox{C. Baltay\unskip,\iYALE}
\mbox{H.R. Band\unskip,\iWISC}
\mbox{M.B. Barakat\unskip,\iLTU}
\mbox{O. Bardon\unskip,\iMIT}
\mbox{T.L. Barklow\unskip,\iSLAC}
\mbox{G.L. Bashindzhagyan\unskip,\iMOSCOW}
\mbox{J.M. Bauer\unskip,\iMISSI}
\mbox{G. Bellodi\unskip,\iOXF}
\mbox{A.C. Benvenuti\unskip,\iBOLO}
\mbox{G.M. Bilei\unskip,\iPERU}
\mbox{D. Bisello\unskip,\iPADO}
\mbox{G. Blaylock\unskip,\iMASS}
\mbox{J.R. Bogart\unskip,\iSLAC}
\mbox{G.R. Bower\unskip,\iSLAC}
\mbox{J.E. Brau\unskip,\iOREG}
\mbox{M. Breidenbach\unskip,\iSLAC}
\mbox{W.M. Bugg\unskip,\iTENN}
\mbox{D. Burke\unskip,\iSLAC}
\mbox{T.H. Burnett\unskip,\iWASH}
\mbox{P.N. Burrows\unskip,\iOXF}
\mbox{R.M. Byrne\unskip,\iMIT}
\mbox{A. Calcaterra\unskip,\iFRAS}
\mbox{D. Calloway\unskip,\iSLAC}
\mbox{B. Camanzi\unskip,\iFERR}
\mbox{M. Carpinelli\unskip,\iPISA}
\mbox{R. Cassell\unskip,\iSLAC}
\mbox{R. Castaldi\unskip,\iPISA}
\mbox{A. Castro\unskip,\iPADO}
\mbox{M. Cavalli-Sforza\unskip,\iUCSC}
\mbox{A. Chou\unskip,\iSLAC}
\mbox{E. Church\unskip,\iWASH}
\mbox{H.O. Cohn\unskip,\iTENN}
\mbox{J.A. Coller\unskip,\iBU}
\mbox{M.R. Convery\unskip,\iSLAC}
\mbox{V. Cook\unskip,\iWASH}
\mbox{R.F. Cowan\unskip,\iMIT}
\mbox{D.G. Coyne\unskip,\iUCSC}
\mbox{G. Crawford\unskip,\iSLAC}
\mbox{C.J.S. Damerell\unskip,\iRAL}
\mbox{M.N. Danielson\unskip,\iCOLO}
\mbox{M. Daoudi\unskip,\iSLAC}
\mbox{N. de Groot\unskip,\iBRI}
\mbox{R. Dell'Orso\unskip,\iPERU}
\mbox{P.J. Dervan\unskip,\iBRUN}
\mbox{R. de Sangro\unskip,\iFRAS}
\mbox{M. Dima\unskip,\iCSU}
\mbox{D.N. Dong\unskip,\iMIT}
\mbox{M. Doser\unskip,\iSLAC}
\mbox{R. Dubois\unskip,\iSLAC}
\mbox{B.I. Eisenstein\unskip,\iILLI}
\mbox{I.Erofeeva\unskip,\iMOSCOW}
\mbox{V. Eschenburg\unskip,\iMISSI}
\mbox{E. Etzion\unskip,\iWISC}
\mbox{S. Fahey\unskip,\iCOLO}
\mbox{D. Falciai\unskip,\iFRAS}
\mbox{C. Fan\unskip,\iCOLO}
\mbox{J.P. Fernandez\unskip,\iUCSC}
\mbox{M.J. Fero\unskip,\iMIT}
\mbox{K. Flood\unskip,\iMASS}
\mbox{R. Frey\unskip,\iOREG}
\mbox{J. Gifford\unskip,\iUVIC}
\mbox{T. Gillman\unskip,\iRAL}
\mbox{G. Gladding\unskip,\iILLI}
\mbox{S. Gonzalez\unskip,\iMIT}
\mbox{E.R. Goodman\unskip,\iCOLO}
\mbox{E.L. Hart\unskip,\iTENN}
\mbox{J.L. Harton\unskip,\iCSU}
\mbox{K. Hasuko\unskip,\iTOHO}
\mbox{S.J. Hedges\unskip,\iBU}
\mbox{S.S. Hertzbach\unskip,\iMASS}
\mbox{M.D. Hildreth\unskip,\iSLAC}
\mbox{J. Huber\unskip,\iOREG}
\mbox{M.E. Huffer\unskip,\iSLAC}
\mbox{E.W. Hughes\unskip,\iSLAC}
\mbox{X. Huynh\unskip,\iSLAC}
\mbox{H. Hwang\unskip,\iOREG}
\mbox{M. Iwasaki\unskip,\iOREG}
\mbox{D.J. Jackson\unskip,\iRAL}
\mbox{P. Jacques\unskip,\iRUTG}
\mbox{J.A. Jaros\unskip,\iSLAC}
\mbox{Z.Y. Jiang\unskip,\iSLAC}
\mbox{A.S. Johnson\unskip,\iSLAC}
\mbox{J.R. Johnson\unskip,\iWISC}
\mbox{R.A. Johnson\unskip,\iCINC}
\mbox{T. Junk\unskip,\iSLAC}
\mbox{R. Kajikawa\unskip,\iNAGO}
\mbox{M. Kalelkar\unskip,\iRUTG}
\mbox{Y. Kamyshkov\unskip,\iTENN}
\mbox{H.J. Kang\unskip,\iRUTG}
\mbox{I. Karliner\unskip,\iILLI}
\mbox{H. Kawahara\unskip,\iSLAC}
\mbox{Y.D. Kim\unskip,\iSOGA}
\mbox{M.E. King\unskip,\iSLAC}
\mbox{R. King\unskip,\iSLAC}
\mbox{R.R. Kofler\unskip,\iMASS}
\mbox{N.M. Krishna\unskip,\iCOLO}
\mbox{R.S. Kroeger\unskip,\iMISSI}
\mbox{M. Langston\unskip,\iOREG}
\mbox{A. Lath\unskip,\iMIT}
\mbox{D.W.G. Leith\unskip,\iSLAC}
\mbox{V. Lia\unskip,\iMIT}
\mbox{C.Lin\unskip,\iMASS}
\mbox{M.X. Liu\unskip,\iYALE}
\mbox{X. Liu\unskip,\iUCSC}
\mbox{M. Loreti\unskip,\iPADO}
\mbox{A. Lu\unskip,\iUCSB}
\mbox{H.L. Lynch\unskip,\iSLAC}
\mbox{J. Ma\unskip,\iWASH}
\mbox{M. Mahjouri\unskip,\iMIT}
\mbox{G. Mancinelli\unskip,\iRUTG}
\mbox{S. Manly\unskip,\iYALE}
\mbox{G. Mantovani\unskip,\iPERU}
\mbox{T.W. Markiewicz\unskip,\iSLAC}
\mbox{T. Maruyama\unskip,\iSLAC}
\mbox{H. Masuda\unskip,\iSLAC}
\mbox{E. Mazzucato\unskip,\iFERR}
\mbox{A.K. McKemey\unskip,\iBRUN}
\mbox{B.T. Meadows\unskip,\iCINC}
\mbox{G. Menegatti\unskip,\iFERR}
\mbox{R. Messner\unskip,\iSLAC}
\mbox{P.M. Mockett\unskip,\iWASH}
\mbox{K.C. Moffeit\unskip,\iSLAC}
\mbox{T.B. Moore\unskip,\iYALE}
\mbox{M.Morii\unskip,\iSLAC}
\mbox{D. Muller\unskip,\iSLAC}
\mbox{V. Murzin\unskip,\iMOSCOW}
\mbox{T. Nagamine\unskip,\iTOHO}
\mbox{S. Narita\unskip,\iTOHO}
\mbox{U. Nauenberg\unskip,\iCOLO}
\mbox{H. Neal\unskip,\iSLAC}
\mbox{M. Nussbaum\unskip,\iCINC}
\mbox{N. Oishi\unskip,\iNAGO}
\mbox{D. Onoprienko\unskip,\iTENN}
\mbox{L.S. Osborne\unskip,\iMIT}
\mbox{R.S. Panvini\unskip,\iVAND}
\mbox{C.H. Park\unskip,\iSOONG}
\mbox{T.J. Pavel\unskip,\iSLAC}
\mbox{I. Peruzzi\unskip,\iFRAS}
\mbox{M. Piccolo\unskip,\iFRAS}
\mbox{L. Piemontese\unskip,\iFERR}
\mbox{K.T. Pitts\unskip,\iOREG}
\mbox{R.J. Plano\unskip,\iRUTG}
\mbox{R. Prepost\unskip,\iWISC}
\mbox{C.Y. Prescott\unskip,\iSLAC}
\mbox{G.D. Punkar\unskip,\iSLAC}
\mbox{J. Quigley\unskip,\iMIT}
\mbox{B.N. Ratcliff\unskip,\iSLAC}
\mbox{T.W. Reeves\unskip,\iVAND}
\mbox{J. Reidy\unskip,\iMISSI}
\mbox{P.L. Reinertsen\unskip,\iUCSC}
\mbox{P.E. Rensing\unskip,\iSLAC}
\mbox{L.S. Rochester\unskip,\iSLAC}
\mbox{P.C. Rowson\unskip,\iCOLU}
\mbox{J.J. Russell\unskip,\iSLAC}
\mbox{O.H. Saxton\unskip,\iSLAC}
\mbox{T. Schalk\unskip,\iUCSC}
\mbox{R.H. Schindler\unskip,\iSLAC}
\mbox{B.A. Schumm\unskip,\iUCSC}
\mbox{J. Schwiening\unskip,\iSLAC}
\mbox{S. Sen\unskip,\iYALE}
\mbox{V.V. Serbo\unskip,\iSLAC}
\mbox{M.H. Shaevitz\unskip,\iCOLU}
\mbox{J.T. Shank\unskip,\iBU}
\mbox{G. Shapiro\unskip,\iLBL}
\mbox{D.J. Sherden\unskip,\iSLAC}
\mbox{K.D. Shmakov\unskip,\iTENN}
\mbox{C. Simopoulos\unskip,\iSLAC}
\mbox{N.B. Sinev\unskip,\iOREG}
\mbox{S.R. Smith\unskip,\iSLAC}
\mbox{M.B. Smy\unskip,\iCSU}
\mbox{J.A. Snyder\unskip,\iYALE}
\mbox{H. Staengle\unskip,\iCSU}
\mbox{A. Stahl\unskip,\iSLAC}
\mbox{P. Stamer\unskip,\iRUTG}
\mbox{H. Steiner\unskip,\iLBL}
\mbox{R. Steiner\unskip,\iADEL}
\mbox{M.G. Strauss\unskip,\iMASS}
\mbox{D. Su\unskip,\iSLAC}
\mbox{F. Suekane\unskip,\iTOHO}
\mbox{A. Sugiyama\unskip,\iNAGO}
\mbox{S. Suzuki\unskip,\iNAGO}
\mbox{M. Swartz\unskip,\iJHU}
\mbox{A. Szumilo\unskip,\iWASH}
\mbox{T. Takahashi\unskip,\iSLAC}
\mbox{F.E. Taylor\unskip,\iMIT}
\mbox{J. Thom\unskip,\iSLAC}
\mbox{E. Torrence\unskip,\iMIT}
\mbox{N.K. Toumbas\unskip,\iSLAC}
\mbox{T. Usher\unskip,\iSLAC}
\mbox{C. Vannini\unskip,\iPISA}
\mbox{J. Va'vra\unskip,\iSLAC}
\mbox{E. Vella\unskip,\iSLAC}
\mbox{J.P. Venuti\unskip,\iVAND}
\mbox{R. Verdier\unskip,\iMIT}
\mbox{P.G. Verdini\unskip,\iPISA}
\mbox{D.L. Wagner\unskip,\iCOLO}
\mbox{S.R. Wagner\unskip,\iSLAC}
\mbox{A.P. Waite\unskip,\iSLAC}
\mbox{S. Walston\unskip,\iOREG}
\mbox{S.J. Watts\unskip,\iBRUN}
\mbox{A.W. Weidemann\unskip,\iTENN}
\mbox{E. R. Weiss\unskip,\iWASH}
\mbox{J.S. Whitaker\unskip,\iBU}
\mbox{S.L. White\unskip,\iTENN}
\mbox{F.J. Wickens\unskip,\iRAL}
\mbox{B. Williams\unskip,\iCOLO}
\mbox{D.C. Williams\unskip,\iMIT}
\mbox{S.H. Williams\unskip,\iSLAC}
\mbox{S. Willocq\unskip,\iMASS}
\mbox{R.J. Wilson\unskip,\iCSU}
\mbox{W.J. Wisniewski\unskip,\iSLAC}
\mbox{J. L. Wittlin\unskip,\iMASS}
\mbox{M. Woods\unskip,\iSLAC}
\mbox{G.B. Word\unskip,\iVAND}
\mbox{T.R. Wright\unskip,\iWISC}
\mbox{J. Wyss\unskip,\iPADO}
\mbox{R.K. Yamamoto\unskip,\iMIT}
\mbox{J.M. Yamartino\unskip,\iMIT}
\mbox{X. Yang\unskip,\iOREG}
\mbox{J. Yashima\unskip,\iTOHO}
\mbox{S.J. Yellin\unskip,\iUCSB}
\mbox{C.C. Young\unskip,\iSLAC}
\mbox{H. Yuta\unskip,\iAOMORI}
\mbox{G. Zapalac\unskip,\iWISC}
\mbox{R.W. Zdarko\unskip,\iSLAC}
\mbox{J. Zhou\unskip.\iOREG}

\it
  \vskip \baselineskip                   
  \centerline{(The SLD Collaboration)}   
  \vskip \baselineskip        
  \baselineskip=.75\baselineskip   
\iADEL
  Adelphi University, Garden City, New York 11530, \break
\iAOMORI
  Aomori University, Aomori , 030 Japan, \break
\iBOLO
  INFN Sezione di Bologna, I-40126, Bologna, Italy, \break
\iBRI
  University of Bristol, Bristol, U.K., \break
\iBRUN
  Brunel University, Uxbridge, Middlesex, UB8 3PH United Kingdom, \break
\iBU
  Boston University, Boston, Massachusetts 02215, \break
\iCINC
  University of Cincinnati, Cincinnati, Ohio 45221, \break
\iCOLO
  University of Colorado, Boulder, Colorado 80309, \break
\iCOLU
  Columbia University, New York, New York 10533, \break
\iCSU
  Colorado State University, Ft. Collins, Colorado 80523, \break
\iFERR
  INFN Sezione di Ferrara and Universita di Ferrara, I-44100 Ferrara, Italy, \break
\iFRAS
  INFN Lab. Nazionali di Frascati, I-00044 Frascati, Italy, \break
\iILLI
  University of Illinois, Urbana, Illinois 61801, \break
\iJHU
  Johns Hopkins University,  Baltimore, Maryland 21218-2686, \break
\iLBL
  Lawrence Berkeley Laboratory, University of California, Berkeley, California 94720, \break
\iLTU
  Louisiana Technical University, Ruston,Louisiana 71272, \break
\iMASS
  University of Massachusetts, Amherst, Massachusetts 01003, \break
\iMISSI
  University of Mississippi, University, Mississippi 38677, \break
\iMIT
  Massachusetts Institute of Technology, Cambridge, Massachusetts 02139, \break
\iMOSCOW
  Institute of Nuclear Physics, Moscow State University, 119899, Moscow Russia, \break
\iNAGO
  Nagoya University, Chikusa-ku, Nagoya, 464 Japan, \break
\iOREG
  University of Oregon, Eugene, Oregon 97403, \break
\iOXF
  Oxford University, Oxford, OX1 3RH, United Kingdom, \break
\iPADO
  INFN Sezione di Padova and Universita di Padova I-35100, Padova, Italy, \break
\iPERU
  INFN Sezione di Perugia and Universita di Perugia, I-06100 Perugia, Italy, \break
\iPISA
  INFN Sezione di Pisa and Universita di Pisa, I-56010 Pisa, Italy, \break
\iRAL
  Rutherford Appleton Laboratory, Chilton, Didcot, Oxon OX11 0QX United Kingdom, \break
\iRUTG
  Rutgers University, Piscataway, New Jersey 08855, \break
\iSLAC
  Stanford Linear Accelerator Center, Stanford University, Stanford, California 94309, \break
\iSOGA
  Sogang University, Seoul, Korea, \break
\iSOONG
  Soongsil University, Seoul, Korea 156-743, \break
\iTENN
  University of Tennessee, Knoxville, Tennessee 37996, \break
\iTOHO
  Tohoku University, Sendai 980, Japan, \break
\iUCSB
  University of California at Santa Barbara, Santa Barbara, California 93106, \break
\iUCSC
  University of California at Santa Cruz, Santa Cruz, California 95064, \break
\iUVIC
  University of Victoria, Victoria, British Columbia, Canada V8W 3P6, \break
\iVAND
  Vanderbilt University, Nashville,Tennessee 37235, \break
\iWASH
  University of Washington, Seattle, Washington 98105, \break
\iWISC
  University of Wisconsin, Madison,Wisconsin 53706, \break
\iYALE
  Yale University, New Haven, Connecticut 06511. \break

\rm
%

\end{center}

\end{document}